\documentclass[12pt]{article}              
\usepackage{amssymb, euscript}
\usepackage{amsmath,bbm}
      
\usepackage{theorem}
\theoremheaderfont{\scshape} 
\theoremstyle{plain}

\newcommand{\haken}{\mathbin{\hbox to 8pt{%
                 \vrule height0.4pt width7pt depth0pt
                 \kern-.4pt
                 \vrule height4pt width0.4pt depth0pt\hss}}}
   
\newcommand{\qcharge}[1]{q_{\scriptscriptstyle{\mathrm{#1}}}}
\newcommand{\PP}{\mathcal{P}}
\newcommand{\wbar}[1]{\overline{#1}}
\newcommand{\seppt}{\setlength\arraycolsep{2pt}}
\newcommand{\be}[3]{\begin{equation}  \label{#1#2#3}}
\newcommand{\bea}[3]{\begin{eqnarray}  \label{#1#2#3}}
\newcommand{\ee}{\end{equation}}
\newcommand{\ba}{\begin{array}}
\newcommand{\ea}{\end{array}}
\newcommand{\eea}{\end{eqnarray}}

\begin{document}

\baselineskip=20pt
\parskip=6pt

%%%%%%%%%%%%%%%%%%%%%%%%%%%%%%%%%%%%%%%%%%%%%%

\thispagestyle{empty}

\vspace{10pt}

\begin{center}{ \LARGE{\bf
On the moduli space of  \\[4mm]
generalized holomorphic maps \\[4mm]
}}

\vspace{35pt}

{\bf Stefano Chiantese}

\vspace{15pt}

{\it  Freie Universit\"at Berlin,\\
FB Mathematik und Informatik,\\
Arnimallee 3, 14195 Berlin, Germany,\\
Email: chiantes@mi.fu-berlin.de.}\\[1mm]

\vspace{8pt}

\vspace{40pt}

{\bf ABSTRACT}

\end{center}

\noindent We compute the anomalies of the topological A and B models 
with target space geometry of Hitchin's generalized type. The dimension of the
moduli space of generalized holomorphic maps is also computed, which turns out
to be equal to the total anomaly if the moduli space is unobstructed. We
obtain this result by identifying the infinitesimal deformations of such maps 
and by using the Grothendieck-Riemann-Roch formula.

\noindent

\vfill

\newpage

%%%%%%%%%%%%%%%%%%%%%%%%%%%%%%%%%%%%%%%%%%%%%%%%%%%%%%

\section{Introduction}

%%%%%%%%%%%%%%%%%%%%%%%%%%%%%%%%%%%%%%%%%%%%%%%%%%

The aim of this paper is to make some concrete preliminary steps towards
an understanding of the moduli space of holomorphic maps from curves to a
generalized Calabi-Yau manifold\footnote{The reason why we mainly focus on
generalized Calabi-Yau manifolds will become clear later on in the paper. 
However, for the topological A and B models the generalized Calabi-Yau 
condition can be properly relaxed as explained in the next paragraph.}. 
Via the localization principle, this has been 
shown by Kapustin~\cite{Kapustin:2003sg} to be the relevant moduli space of 
the topological A and B models with Hitchin's generalized complex
geometry~\cite{Hitchin:2002}. Such maps generalize the notion of
holomorphic embeddings of curves into a Calabi-Yau manifold. These give  
the moduli space over which correlators of the topological A model on a  
Calabi-Yau manifold localize. In this case it is well known that coupling the
A model to worldsheet gravity is of fundamental importance. In fact, there are
no genus $g>1$ correlators for a fixed worldsheet geometry. This can be 
seen by selection rules which come from the anomalies in the twisted theory. 
From a mathematical point of view this is due to the fact that there are no 
holomorphic maps from a genus $g>1$ Riemann surface to a Calabi-Yau manifold.  

These considerations tell us that for computing 
correlators of topological string theory with generalized complex geometry, 
one has to tackle first the problem of the anomalies in the twisted theory. 
In this paper we compute the vector and axial anomalies of the A and B models
on a generalized K\"ahler manifold~\cite{Gualtieri:2004}, which has to 
satisfy a geometry constraint so that the twist can be 
performed~\cite{Kapustin:2004gv}. The key facts one has to take into account 
is that left and right fermions are defined in terms of two different complex 
structures of the target space geometry, and the Dirac operator in the kinetic
terms of the untwisted theory becomes the Dolbeault operator as a result of 
the twist. As a consequence, the anomalies are given in terms of the indices 
of the Dolbeault operator coupled to the holomorphic tangent spaces of the two 
complex structures. 

We also compute the dimension of the moduli space of holomorphic maps from 
curves to a generalized K\"ahler manifold. To this end one has to identify the
deformations of such maps. We find that the deformations (obstructions) lie 
in the zero (first) \v Cech cohomology of sections of the maximal 
isotropic subbundle of a generalized complex structure. If the moduli space is 
unobstructed, its virtual or expected dimension coincides with the total 
anomaly when the above geometry constraint is taken into account. 
Moreover, we find that if there are no marked points on the Riemann surface, 
generalized Calabi-Yau threefolds assume a special role. 
In fact, the virtual dimension becomes zero regardless 
the genus of the Riemann surface. In the final part of the paper we use this
result to speculate on a possible generalization of Gromow-Witten theory. 

%%%%%%%%%%%%%%%%%%%%%%%%%%%%%%%%%%%%%%%%%%%%%%%%%%%%%%

\section{The anomalies of the twisted theories}

%%%%%%%%%%%%%%%%%%%%%%%%%%%%%%%%%%%%%%%%%%%%%%%%%%

We want to compute the anomalies of the A and B models when the 
target space is a generalized K\"ahler manifold. 
To this end, it is useful to recall the case
of the A model with its usual target space
geometry which is K\"ahler.\footnote{For simplicity, we do not consider the 
Landau-Ginzburg model. In this case, the $\sigma$-model can
be twisted assigning zero vector charge to all chiral superfields.} 
In such a case, while we do not need to impose an anomaly cancellation 
condition to define the model, the presence of an axial anomaly in the 
twisted $\sigma$-model leads to a selection rule for the correlation 
functions.\footnote{There is also
a selection rule coming from the vector symmetry. Here, we are principally
interested in the axial anomaly since it is related to the dimension of the
moduli space of holomorphic maps. The situation will change for generalized
geometries as we will see below.} The computation of this anomaly is closely
related to that of the axial anomaly of the untwisted $\sigma$-model, 
where it is given by the index of the Dirac operator. 
The main difference is that the Dirac operator becomes the Dolbeault operator 
as a result of the twist. The anomaly is then given by the index of the
Dolbeault operator, which is related to the dimension of the moduli space of
holomorphic maps via the Grothendieck-Riemann-Roch formula (GRR). 

The main fact one has to take into account to extend these 
considerations to topological models with Hitchin's generalized geometries 
is that left and right moving fermions are defined in terms of two different 
complex structures of the target space. As a result, one is forced to cancel a 
vector anomaly to perform the A twist~\cite{Kapustin:2004gv}, while for the 
usual K\"ahler geometry the vector symmetry is never anomalous. The 
considerations of the precedent paragraph then suggest us that the topological 
models should have axial and vector anomalies.

To compute these anomalies we need to know the kinetic terms of the fields and 
their R-charges. The anomaly is given by a mismatch of the fermion zero
modes, which is expressed in terms of the indeces of the operators appearing 
in the kinetic terms. These indices are then computed by the index theorem.  
The R-charges are given by the following table
\begin{equation}\label{twistedcharges}\mbox{
\begin{tabular}{l|rr}
                     & $\qcharge{V}$ & $\qcharge{A}$ \\\hline
  $\PP_+\psi_+$        & $-1$  & $-1$ \\
  $\wbar{\PP}_+\psi_+$ & $+1$  & $+1$ \\
  $\PP_-\psi_-$        & $-1$  & $+1$ \\
  $\wbar{\PP}_-\psi_-$ & $+1$  & $-1$ \\ 
\end{tabular}}\end{equation}
where 
\begin{displaymath}
\PP_{\pm} = \frac{1}{2}(1-iI_{\pm})
\end{displaymath}
are the projectors $\PP_{\pm} : T \otimes \mathbb C \to T^{(1,0)}_{\pm}$. The
kinetic terms are 
\begin{equation}
\label{kterms}
g(\psi_+, D_z \PP_+ \psi_+) = g_{i \bar j} \psi^{\bar j}_+ D_z \psi^i_+ 
\,, \qquad
g(\psi_-, D_{\bar z} \PP_- \psi_-) = g_{i \bar j} \psi^{\bar j}_- 
D_{\bar z} \psi^i_- \,.
\end{equation}
Note the slight abuse of notation in the right hand sides of the above 
equations as the decomposition in holomorphic and antiholomorphic 
components of $T \otimes \mathbb C$ is not unique. 
The table~(\ref{twistedcharges}) and the kinetic 
terms~(\ref{kterms}) yield the fermion zero mode mismatches
{\seppt
\begin{eqnarray}
\mathrm{Vector \;\: (V)} \, &:& \quad - \mathrm{dim\, Ker} D^+_z + 
\mathrm{dim Ker} D^+_{\bar z} 
- \mathrm{dim\, Ker} D^-_{\bar z} + \mathrm{dim\, Ker} D^-_z \,, 
\nonumber \\[2pt]
\mathrm{Axial \;\: (A)} \, &:& \quad - \mathrm{dim\, Ker} D^+_z + 
\mathrm{dim\, Ker} D^+_{\bar z} 
+ \mathrm{dim\, Ker} D^-_{\bar z} - \mathrm{dim\, Ker} D^-_z \,. \nonumber
\end{eqnarray}}
The superscript $+$ or $-$  keeps track of the tangent bundle to which 
the operators couple. Before performing the twist, these operators are
Dirac operators, and we have the anomalies
\begin{equation}
\label{anomalyva}
(\mathrm{Anomaly})_{\mathrm{V/A}} = \nu^+(D) \mp \nu^-(D) \,,
\end{equation}
where $\nu^{\pm}(D)$ are the indeces of the Dirac operators coupled to the
tangent bundles $T^{(1,0)}_{\pm}$. These indeces are given by the index
theorem,
\begin{displaymath}
\nu^{\pm}(D) = \int_{\Sigma} \phi^* c_1(T^{(1,0)}_{\pm}) \,,
\end{displaymath}
where $\phi$ is a map from a Riemann surface $\Sigma$ to 
the target space $X$. Therefore, we find 
\begin{equation}
\label{untwva}
(\mathrm{Anomaly})_{\mathrm{V/A}} = \int_{\Sigma} \phi^* 
\big\{c_1(T^{(1,0)}_+) \mp c_1(T^{(1,0)}_-)\big\} =  
\int_{\Sigma} \phi^* c_1(L_{2/1}) \,.
\end{equation}
$L_{1/2}$ are the maximal isotropic subbundles of $(T \oplus T^*) \otimes
\mathbb C$ coming from the generalized complex structures 
\begin{displaymath}
{\cal J}^b_{1/2}= e^b \, \frac{1}{2} 
\begin{pmatrix} 
           I_+ \pm I_- &  -(\omega_+^{-1} \mp \omega_-^{-1}) \\ 
\omega_+ \mp \omega_-  &  -(I_+^* \pm I_-^*) 
\end{pmatrix} e^{-b} \,, \qquad e^b = \begin{pmatrix}
 1 & \;\; \\
 b & \;\; 1
\end{pmatrix} \,, 
\end{displaymath} 
which are given by
{\seppt
\begin{displaymath}
\begin{array}{rcl} 
\label{435}
L_1 & = & L^+_1 \oplus L^-_1 = \big\{v +(b+g)v: v \in T_+^{(1,0)} \} \oplus 
\{v + (b-g)v : v \in T_-^{(1,0)} \big\} \,, \\[5pt] 
L_2 & = & L^+_2 \oplus L^-_2 = \big\{v +(b+g)v : v \in T_+^{(1,0)} \} \oplus 
\{v +(b-g)v : v \in T_-^{(0,1)} \big\} \,.
\end{array}
\end{displaymath}}
The equations~(\ref{untwva}) have been found by Kapustin and 
Li~\cite{Kapustin:2004gv}. They show the existence of a vector anomaly in
addition to the axial anomaly, and imply that the geometry of the 
A/B model is constrained via $c_1(L_{2/1}) =0$. When $I_+=I_-$, one finds
the known fact that the B model can only be defined 
on a Calabi-Yau manifold while no condition is required to define 
the A model.  

The A and B models have vector and axial symmetries.
These symmetry are spoiled at the quantum level by
the presence of an anomaly, which is given in terms of the indeces of the
Dolbeault operators. In fact, the equations~(\ref{anomalyva}) become
\begin{displaymath}
(\mathrm{Anomaly})^{\mathrm{tw}}_{\mathrm{V/A}} = \nu^+(\bar \partial) 
\mp \nu^-(\bar \partial) \,.
\end{displaymath}
The Dirac operator has become a Dolbeault operator as a result of the twist,
and the superscript tw indicates that we are computing an anomaly for 
a twisted theory. 
The index theorem asserts that 
\begin{displaymath}
\nu^{\pm}(\bar \partial) = \int_{\Sigma} \mathrm{ch} (\phi^* T^{(1,0)}_{\pm}) 
\mathrm{td}(T_{\Sigma}) = \mathrm{dim}_{\mathbb C} X (1-g) +
\int_{\Sigma} \phi^* c_1 (T^{(1,0)}_{\pm}) 
\end{displaymath}
for a Riemann surface $\Sigma$ of genus $g$.
We have used that in the expansion of the Chern character, only forms of 
degree less than or equal to two enter since the bundles 
$\phi^* T^{(1,0)}_{\pm}$ are over $\Sigma$. Moreover, $\phi^* T^{(1,0)}_{\pm}$
have rank equal to $\mathrm{dim}_{\mathbb C} X$. The anomalies take the form
{\seppt
\begin{eqnarray}
(\mathrm{Anomaly})^{\mathrm{tw}}_{\mathrm V} &=&
\int_{\Sigma} \phi^* \big\{c_1(T^{(1,0)}_+) - c_1(T^{(1,0)}_-)\big\} =  
\int_{\Sigma} \phi^* c_1(L_2) \,, \nonumber \\[2pt]
(\mathrm{Anomaly})^{\mathrm{tw}}_{\mathrm A} &=&
2 \mathrm{dim}_{\mathbb C} X (1-g) +
\int_{\Sigma} \phi^* \big\{c_1(T^{(1,0)}_+) + c_1(T^{(1,0)}_-)\big\} 
\nonumber \\[2pt] &=& 2 \mathrm{dim}_{\mathbb C} X (1-g) + 
\int_{\Sigma} \phi^* c_1(L_1) \,. \nonumber
\end{eqnarray}}
Note that the vector anomaly is present only in the B model with 
$c_1(L_2) \neq 0$. 
Taking into account the vector and axial R-charges of the operators in the 
correlation functions, these anomalies lead to selection rules. 
The total anomaly is
{\seppt
\begin{eqnarray}
(\mathrm{Total\: anomaly})^{\mathrm{tw}} &=& 
(\mathrm{Anomaly})^{\mathrm{tw}}_{\mathrm V} + 
(\mathrm{Anomaly})^{\mathrm{tw}}_{\mathrm A} \nonumber \\[2pt]
&=& 2 \mathrm{dim}_{\mathbb C} X (1-g) +
\int_{\Sigma} \phi^* \big\{c_1(L_1) + c_1(L_2)\big\} \nonumber \,.
\end{eqnarray}}
Imposing the geometry constraints, we get 
\begin{equation}
\label{totanab}
(\mathrm{Total\: anomaly})^{\mathrm{tw}}_{\mathrm{A/B}} = 
2 \mathrm{dim}_{\mathbb C} X (1-g) +
\int_{\Sigma} \phi^* c_1(L_{1/2}) 
\end{equation}
for the A and B models respectively. We shall see that these equations 
are related to the dimension of the moduli spaces of generalized holomorphic
maps.  

%%%%%%%%%%%%%%%%%%%%%%%%%%%%%%%%%%%%%%%%%%%%%%%%%%%%%%

\section{The dimension of the moduli space}

%%%%%%%%%%%%%%%%%%%%%%%%%%%%%%%%%%%%%%%%%%%%%%%%%%

We want to compute the dimension of the moduli space of generalized
holomorphic maps by using the GRR formula. First, recall that the instantons 
of the A and B models are given 
by~\cite{Kapustin:2003sg,Kapustin:2004gv}
\begin{displaymath}
\frac{1}{2}(1 - i{\cal J}^b_{2/1})
\begin{pmatrix} 
i \partial_2 \phi \\
g \partial_1 \phi \end{pmatrix} = 0 \,.
\end{displaymath}
This can be rewritten as 
\begin{displaymath}
{\cal J}^b_{1/2} (i \circ d \phi) = (i \circ d \phi) I_{\Sigma} \,,
\end{displaymath}
where $ I_{\Sigma}$ is the complex structure of the Riemann surface, and  
\begin{displaymath}
\setlength{\unitlength}{1cm}
\begin{picture}(5,1) 
\put(2.6,0.4){$i$}
\put(0.8,0.4){\footnotesize{$d \phi$}}
\put(0,0){$T_{\Sigma} \; \longrightarrow \; T_X \; \hookrightarrow 
\; T_X \oplus T^*_X \,.$}
\end{picture}
\end{displaymath}
One can use the projection operators 
\begin{displaymath}
\PP_{L_{1/2}} = \frac{1}{2}(1-i\mathcal J_{1/2}) \,, \qquad 
\PP_{\Sigma} = \frac{1}{2}(1-i I_{\Sigma}) \,,
\end{displaymath}
to recast the above equation into
\begin{equation}
\label{proinst}
\PP_{L_{1/2}} (i \circ d \phi) = (i \circ d \phi) \PP_{\Sigma} \,.
\end{equation}
Since $(T_X \oplus T^*_X)\otimes \mathbb C = L_{1/2} \oplus \overline L_{1/2}$,
we have 
\begin{displaymath}
(i \circ d \phi) = (i \circ d \phi)_{L_{1/2}} + 
(i \circ d \phi)_{\overline L_{1/2}} \,,
\end{displaymath}
with $(i \circ d \phi)_{L_{1/2}}$ and 
$(i \circ d \phi)_{\overline L_{1/2}}$ sections of $L_{1/2}$ and 
$\overline L_{1/2}$. 
Substituting the last equation into~(\ref{proinst}),  we get
\begin{displaymath}
(i \circ \bar \partial \phi)_{L_{1/2}} =
(i \circ \partial \phi)_{\overline L_{1/2}} \,.
\end{displaymath}
These are sections of orthogonal bundles. The only possibility is
\begin{displaymath}
(i \circ \bar \partial \phi)_{L_{1/2}} = 0 = 
(i \circ \partial \phi)_{\overline L_{1/2}} \,.
\end{displaymath}
This indicates that infinitesimal deformations of generalized holomorphic maps 
are given in terms of $\bar \partial$-closed section of the bundle
$\phi^* L_{1/2}$, \emph{i.e.} they are elements of
\begin{displaymath}
H^{0,0}_{\bar \partial}(\Sigma,\phi^* L_{1/2}) = 
\check H^0(\Sigma,\phi^*L_{1/2}) \,.
\end{displaymath}
Therefore, the dimension
of the moduli space $ \mathcal M^{\small{\mathrm{A/B}}}_g$ of generalized 
holomorphic maps from a Riemann surface of genus $g$ to the target space
of the A/B model is given by
\begin{displaymath}
\mathrm{dim} \mathcal M^{\small{\mathrm{A/B}}}_g =   
\mathrm{dim} \check H^0(\Sigma,\phi^*L_{1/2}) \,.
\end{displaymath}
By the GRR formula,
\begin{displaymath}
\mathrm{dim} \check H^0(\Sigma,\phi^* L_{1/2}) - 
\mathrm{dim} \check H^1(\Sigma,\phi^* L_{1/2}) = 
\int_{\Sigma} \mathrm{ch} (\phi^* L_{1/2}) 
\mathrm{td}(T_{\Sigma}) \,.
\end{displaymath}
Considering that $(T_X \oplus T^*_X) \otimes \mathbb C = 
L_1 \oplus \overline L_2$, we get
\begin{displaymath}
\mathrm{dim} \mathcal M^{\small{\mathrm{A/B}}}_g = 
\mathrm{dim} \check H^1(\Sigma,\phi^* L_{1/2}) +
2 \mathrm{dim}_{\mathbb C} X (1-g) + \int_{\Sigma} 
\phi^* c_1 (L_{1/2}) \,.
\end{displaymath}
If the moduli space of generalized holomorphic maps is unobstructed, 
\emph{i.e.}
\begin{displaymath}
\check H^1(\Sigma,\phi^* L_{1/2}) = 0 \,,
\end{displaymath}
we get
\begin{displaymath}
\mathrm{dim} \mathcal M^{\small{\mathrm{A/B}}}_g = 
2 \mathrm{dim}_{\mathbb C} X (1-g) + \int_{\Sigma} 
\phi^* c_1 (L_{1/2}) \,.
\end{displaymath}
Comparing this equation with~(\ref{totanab}), we have shown that in the
unobstructed case the total anomaly coincides with the dimension of the 
moduli space of generalized holomorphic maps. 

In this paper we will not investigate criteria under which the unobstructedness
assumption holds. Instead, we simply assume from now on that under suitable 
conditions the moduli space of generalized holomorphic maps is unobstructed. 
Then, the equations above imply the triviality at $g>1$ of the partition 
function of topological models defined on a generalized Calabi-Yau
manifold, for which $c_1 (L_1)=0=c_1 (L_2)$.  The partition function becomes
non-trivial by coupling to worldsheet gravity. Consequently, we also have to 
integrate over the moduli space $\overline{\mathcal M}_{g,n}$ of 
curves with $n$ marked points. This enhances the dimension of the
moduli space, yielding the so-called virtual dimension. 
We indicate by $\overline{\mathcal M}^{\small{\mathrm{A/B}}}_{g,n}$
the moduli space of generalized holomorphic maps from Riemann surfaces with
$n$ marked points to a generalized K\"ahler manifold with $c_1 (L_{2/1})=0$.
From a physical viewpoint, $n$ is the number of operator 
insertions. In other words, an $n$-point correlation function of 
topological string theory is obtained by integrating over 
$\overline{\mathcal M}^{\small{\mathrm{A/B}}}_{g,n}$. Since 
\begin{displaymath}
\mathrm{dim}_{\mathbb C} \mathcal M_{g,n} = 3(g-1) + n \,,
\end{displaymath}
the virtual dimension is
\begin{displaymath}
\mathrm{dim} \overline{\mathcal M}^{\small{\mathrm{A/B}}}_{g,n}  = 
2 (\mathrm{dim}_{\mathbb C} X - 3)(1-g) + 2n + \int_{\Sigma} 
\phi^* c_1 (L_{1/2}) \,.
\end{displaymath}
For generalized Calabi-Yau threefolds of complex dimension three 
\begin{displaymath}
\mathrm{dim} \overline{\mathcal M}^{\small{\mathrm{A/B}}}_{g,0}  = 0 \,.
\end{displaymath}
This suggests that the computation of the partition function reduces 
to a counting problem. If we consider the topological A model on a 
generalized Calabi-Yau manifold and take $I_+=I_-$, we get the A model on a 
Calabi-Yau manifold, where the counting problem is know to lead to the
Gromow-Witten invariants. They are symplectic invariants of the underlying 
Calabi-Yau manifold. It is conceivable that the counting problem in the 
generalized case might lead to new invariants. The invariants of the
A model on a generalized Calabi-Yau manifold will degenerate to the 
Gromow-Witten invariants in the extreme case of $I_+=I_-$. But in the case 
$I_+ \neq I_-$ such invariants will be of generalized type, which interpolates 
between those of symplectic and complex type.

%%%%%%%%%%%%%%%%%%%%%%%%%%%%%%%%%%%%%%%%%%%%%%%%%%%%%%

\section{Conclusions and outlook}

%%%%%%%%%%%%%%%%%%%%%%%%%%%%%%%%%%%%%%%%%%%%%%%%%%

An important issue that deserves to be studied is to find the criteria
under which the moduli space is unobstructed. In fact, we have seen that in
such a case generalized Calabi-Yau threefolds assume a special role. One 
should note that for holomorphic maps into a Calabi-Yau threefold, there are 
often obstructions that yield a positive virtual dimension.

In the unobstructed case of holomorphic maps from
curves with no marked points to a Calabi-Yau threefold, the virtual dimension
is zero. Therefore, the moduli space is given by a number of points, and the 
computation of the partition function reduces to a
counting point problem that is known to lead to Gromov-Witten invariants. We
have seen above that in the Hitchin's generalized setup under suitable
conditions the virtual dimension is zero. This indicates that also in this
case the computation of the partition function reduces to a counting point
problem that could lead to a generalization of Gromov-Witten theory. This idea
is supported by the physics fact that the A model on a generalized Calabi-Yau
threefold with the two complex structures identified is nothing but the
A model on a Calabi-Yau threefold. Therefore, the generalized
Gromov-Witten theory, which might exist when the two complex structures are
different, has to degenerate to the usual one when the two structures are 
equal.  

The computation of the partition function is highly non trivial. Therefore,
one should start with genus zero correlators. For usual genus zero holomorphic 
maps, selection rules assert that only the three-point correlator of 
three $(1,1)$-forms is non zero. This
correlator leads to the deformed intersection theory. In other words, the
intersection number of three divisors of a Calabi-Yau threefold is deformed
by the quantum cohomology ring of the target space, which contains genus zero
Gromov-Witten invariants. In the generalized setup the situation should be
quite similar. However, the $(1,1)$-forms should be replaced by Hitchin's 
polyforms of the target space, and one should ask first what the 
intersection number of polycycles could be. The next step would consist in
seeing how the quantum cohomology ring deforms this generalized intersection 
number.

%%%%%%%%%%%%%%%%%%%%%%%%%%%%%%%%%%%%%%%%%%%%%%%%%%%%%%

\section{Acknowledgments}

%%%%%%%%%%%%%%%%%%%%%%%%%%%%%%%%%%%%%%%%%%%%%%%%%%

The author would like to thank Albrecht Klemm and Stefan Theisen for 
interesting discussions, and Florian Gmeiner and Frederik Witt for
proofreading the paper. This work was supported by SFB 647.

%%%%%%%%%%%%%%%%   Bibliography  %%%%%%%%%%%%%%%%%%%%%%%%%%%%%%%%
\bibliographystyle{JHEP}
\addcontentsline{toc}{chapter}{Bibliography}
\bibliography{references}
%%%%%%%%%%%%%%%%%%%%%%%%%%%%%%%%%%%%%%%%%%%%%%%%%%%%%%%%%%%%%%%%%

\end{document}